\documentclass[aps,prx,twocolumn,superscriptaddress,notitlepage,nopacs,amsmath,amstex,amssymb,citeautoscript,longbibliography,floatfix,letter,twocolumn]{revtex4-2}

\usepackage{enumitem}
\usepackage{natbib}
\usepackage[english]{babel}
\usepackage{letltxmacro}
\usepackage{latexsym}
\LetLtxMacro{\ORIGselectlanguage}{\selectlanguage}
\makeatletter
\DeclareRobustCommand{\selectlanguage}[1]{%
  \@ifundefined{alias@\string#1}
    {\ORIGselectlanguage{#1}}
    {\begingroup\edef\x{\endgroup
       \noexpand\ORIGselectlanguage{\@nameuse{alias@#1}}}\x}%
}
\newcommand{\definelanguagealias}[2]{%
  \@namedef{alias@#1}{#2}%
}
\makeatother
\definelanguagealias{en}{english}
\definelanguagealias{English}{english}

\newcommand{\be}{\begin{equation}}
\newcommand{\ee}{\end{equation}}
\newcommand{\bea}{\begin{eqnarray}}
\newcommand{\eea}{\end{eqnarray}}

\usepackage{amsthm}

\usepackage{graphicx}
\usepackage{bm}
\usepackage{physics}
\usepackage[dvipsnames]{xcolor}
\usepackage{algorithm}
\usepackage{algpseudocode}
\usepackage{blindtext}
\usepackage{orcidlink}
\usepackage[normalem]{ulem}
\makeatletter
\newcommand{\printfnsymbol}[1]{%
  \textsuperscript{\@fnsymbol{#1}}%
}
\makeatother
\usepackage{hyperref}
\usepackage{soul}
\hypersetup{
    unicode=false,          
    pdftoolbar=false,        
    pdfmenubar=true,        
    pdffitwindow=false,     
    pdfstartview={FitH},    
    pdftitle={},    
    pdfauthor={Stefan H. Sack and Daniel J. Egger},     
    pdfsubject={},   
    pdfcreator={},   
    pdfproducer={}, 
    pdfkeywords={}, 
    pdfnewwindow=true,      
    colorlinks=true,       
    linkcolor=blue,          
    citecolor=blue,        
    filecolor=blue,      
    urlcolor=blue           
}

\def\dout{\bgroup
 \markoverwith{\lower-0.2ex\hbox
 {\kern-.03em\vbox{\hrule width.2em\kern0.45ex\hrule}\kern-.03em}}%
 \ULon}
\MakeRobust\dout

\begin{document}
\title{Large-scale quantum approximate optimization on non-planar graphs with machine learning noise mitigation}

\author{Stefan H. Sack \orcidlink{0000-0001-5400-8508}}
\affiliation{Institute of Science and Technology Austria (ISTA), Am Campus 1, 3400 Klosterneuburg, Austria}
\email{stefan.sack@ista.ac.at}

\author{Daniel J. Egger \orcidlink{0000-0002-5523-9807}}
\affiliation{IBM Quantum, IBM Research Europe -- Zurich, S\"aumerstrasse 4, CH-8803 R\"uschlikon, Switzerland}
\email{deg@zurich.ibm.com}

\date{\today}
\begin{abstract}
Quantum computers are increasing in size and quality, but are still very noisy. Error mitigation extends the size of the quantum circuits that noisy devices can meaningfully execute. 
However, state-of-the-art error mitigation methods are hard to implement and the limited qubit connectivity in superconducting qubit devices restricts most applications to the hardware's native topology. 
Here we show a quantum approximate optimization algorithm (QAOA) on non-planar random regular graphs with up to 40 nodes enabled by a machine learning-based error mitigation. 
We use a swap network with careful decision-variable-to-qubit mapping and a feed-forward neural network to optimize a depth-two QAOA on up to 40 qubits. 
We observe a meaningful parameter optimization for the largest graph which requires running quantum circuits with 958 two-qubit gates.
Our work emphasizes the need to mitigate samples, and not only expectation values, in quantum approximate optimization. 
These results are a step towards executing quantum approximate optimization at a scale that is not classically simulable.
Reaching such system sizes is key to properly understanding the true potential of heuristic algorithms like QAOA.
\end{abstract}

\maketitle

\section{Introduction}

Quantum information processing holds the promise of drastically speeding up a number of interesting computational tasks ranging from optimization~\cite{Abbas2023}, chemistry~\cite{McArdle2020}, finance~\cite{Egger2020}, machine learning~\cite{Biamonte2017, Benedetti2019}, and high-energy physics~\cite{Guan2020, Dimeglio2023}.
The size and quality of noisy quantum computers is progressing rapidly. 
In particular, error mitigation tools~\cite{Temme2017, Endo2018}, e.g., zero-noise extrapolation (ZNE)~\cite{Li2017, Temme2017} and probabilistic error cancellation (PEC)~\cite{Temme2017}, extend the reach of noisy hardware~\cite{Kandala2018}.

In ZNE multiple logically equivalent copies of a circuit are run under different noise amplification factors~$c$.
Based on the noisy results, extrapolation to the zero-noise limit produces a biased estimation of the noiseless expectation value.
ZNE can be performed with pulses which results in small stretch factors $c$ close to one~\cite{Kandala2018}.
However, pulse-based ZNE is almost impossible for users of a cloud-based quantum computer to implement due to the onerous calibration.
As alternative, digital ZNE folds gates such as CNOTs which produce large stretch factors $c=2m+1$ where $m$ is the number of times a gate is folded~\cite{Giurgica2020}.
If the original circuit is deep compared to the noise then the first fold $m=1$ results in noise rendering the extrapolation useless.
Partial folding prevents this by folding a sub-set of the gates in a circuit~\cite{LaRose2022}.

PEC learns a sparse model of the noise~\cite{vandenBerg2023}.
The non-physical inverse of the noise channel is applied through a quasi-probability distribution to recover an unbiased expectation value.
However, the large shot overhead of PEC can be prohibitive.
This is why large experiments resort to Probabilistic Error Amplification (PEA), a form of ZNE in which the learnt error channels are amplified~\cite{Kim2023}.
PEA avoids pulse calibration but requires an onerous noise learning like PEC.

Pulse-based approaches can also error mitigate variational algorithms enabled by, e.g., open-pulse~\cite{Alexander2020}.
Scaled cross-resonance gates~\cite{Stenger2021, Earnest2021} can implement ZNE~\cite{Vazquez2022} and reduce the schedule duration without calibrating pulses~\cite{weidenfeller2022scaling}.
Other approaches, inspired by optimal control~\cite{Magann2021}, leave it up to the classical optimizer to shape the pulses resulting in shorter schedules~\cite{Meitei2021, Egger2023}.

Supervised learning benefits a wide range of scientific fields, including quantum physics~\cite{Carleo2017}. 
In particular, it can mitigate hardware noise in quantum computations. 
Kim \emph{et al.}~\cite{kim2020quantum} adjust the probabilities estimated from measurements of quantum circuits with neural networks.
They show an effective reduction in errors with a method that scales exponentially with system size.
Czarnik \emph{et al.}~\cite{czarnik2021error} propose a scalable method to error mitigate observables, rather than the full state vector, with linear regression.
They efficiently generate training data by computing expectation values of Clifford circuits on noiseless simulators and noisy quantum hardware.
Similarly, Strikis~\emph{et al.}~\cite{strikis2021learning-based} present a method that learns noise mitigation from Clifford data.
They error mitigate a quantum circuit by simulating multiple versions of it in which non-Clifford gates are replaced with gates that are efficient to simulate classically.
These methods successfully mitigate noise on both real quantum hardware and simulations of imperfect quantum computers.

Combinatorial problems are regularly encountered in practical settings such as finance and vehicle routing.
The quantum approximate optimization algorithm (QAOA)~\cite{farhi2014quantum} may help solve such problems by mapping the cost function to a spin Hamiltonian and finding its ground state with a suitable variational Ansatz~\cite{Lucas2014}.
Crucially, many problems of practical interest are non-planar~\cite{Markowitz1952}, but common superconducting qubit architectures have a grid~\cite{Harrigan2021} or heavy-hexagonal~\cite{Chamberland2020} coupling map.
Recently, QAOA experiments with a connectivity matching the hardware coupling map have been reported for 27~\cite{weidenfeller2022scaling} and 127~\cite{Pelofske2023} qubits with up to QAOA depth-two.
By contrast to industry relevant problems, these instances are very sparse.
Moreover, classical solvers perform well, especially on sparse problems~\cite{Rehfeldt2023}.
While brute-force classical simulation methods of quantum circuits can handle up to around 50 qubits~\cite{Pan2022, Kissinger2022} tensor-product-based methods are capable of simulating much larger QAOA circuits.
For example, Lykov \emph{et al.}~\cite{Lykov2022} report simulating a single depth-one QAOA amplitude with up to 210 qubits and 1785 gates on a supercomputer.
There is therefore a dire need to implement denser and larger problems than those in current demonstrations on hardware.

In this work, we make two contributions.
Inspired by Refs.~\cite{czarnik2021error, strikis2021learning-based}, we present an error mitigation strategy based on a neural network that uses measurements of noisy observables and compares them to their ideal values.
Second, we go one step beyond the hardware-native topology by implementing in hardware random three regular graphs with up to forty nodes.
We achieve this by combining swap networks~\cite{weidenfeller2022scaling, Harrigan2021}, and the SAT-based initial mapping of Matsuo \emph{et al.}~\cite{Matsuo2022} which was so far only numerically studied.

This paper is structured as follows.
In Sec.~\ref{sec:qaoa} we introduce the QAOA and discuss its implementation on hardware.
Sec.~\ref{sec:ml_qem} discusses machine-learning assisted quantum error mitigation.
In Sec.~\ref{sec:ml_qaoa} we combine the QAOA implementation advances of Sec.~\ref{sec:qaoa} and the error mitigation approach of Sec.~\ref{sec:ml_qem} to train depth-two QAOA circuits on hardware.
We discuss our results and conclude in Sec.~\ref{sec:conclusion}.

\section{Quantum Approximate Optimization Algorithm\label{sec:qaoa}}
 
The QAOA was initially developed to solve the maximum cut (\textsc{MaxCut}) problem~\cite{farhi2014quantum}, but it also applies to any Quadratic Unconstrained Binary Optimization (QUBO) as exemplified by Refs.~\cite{Streif2021, farhi2020needs, Marwaha2021}.
\textsc{MaxCut} requires cutting the set of nodes $V$ of a given undirected graph $G=(V, E)$ into two groups to maximize the number of edges in $E$ traversed by the cut.
This problem, as many others, is equivalent to finding the ground state of an Ising Hamiltonian for an $n$-qubit system, where $n=|V|$ is the number of decision variables~\cite{Lucas2014}.

A depth-$p$ QAOA for an unweighted \textsc{MaxCut} minimizes the expectation value of the cost function Hamiltonian $H_C = \sum_{(i,j) \epsilon E} \sigma^z_i \sigma^z_j$ under the variational state
\begin{equation}\label{eq:QAOA_ansatz}
\ket{\bm{\beta}, \bm{\gamma}} = \prod_{i=1}^{p} e^{-i \beta_i H_B} e^{-i \gamma_i H_C} \ket{+}^{\otimes n}.
\end{equation}
The initial product state $\ket{+}^{\otimes n}$ is an equal superposition of all possible solutions.
It is also the ground state of the mixer Hamiltonian $H_B = -\sum_i^n \sigma^x_i$~\cite{farhi2014quantum}.
The circuit depth, controlled by $p$, determines the number of applications of the Hamiltonians.
A classical optimizer varies the angles $\bm{\beta}=(\beta_1,\ldots,\beta_p)$ and $\bm{\gamma}=(\gamma_1,\ldots,\gamma_p)$ to minimize the energy expectation value $E(\bm{\beta}, \bm{\gamma})=\langle H_C \rangle$ in a closed-loop with the quantum computer until the parameters $\bm{\beta}, \bm{\gamma}$ converge.
We denote the optimized parameters by $\bm{\theta}^\star=(\bm{\beta}^\star, \bm{\gamma}^\star) = \arg \min_{\bm{\beta}, \bm{\gamma}} E(\bm{\beta}, \bm{\gamma})$.

\subsection{Implementation on superconducting hardware}

In hardware, $e^{-i\beta_i H_B}$ is trivially implemented by single-qubit $\text{R}_X$ rotations applied to all qubits.
The cost-operator, however, creates a network of $\text{R}_{ZZ}$ gates that matches the graph connectivity.
Noisy quantum hardware can run graphs with many nodes if their topology matches the connectivity of the qubits~\cite{Pelofske2023}.
However, SWAP gates must be inserted in the circuit when the structure of $G$ does not match the native coupling map between the qubits.
This severely limits the number of nodes that can be considered~\cite{Franca2020, weidenfeller2022scaling}.

Transpiler passes are responsible for routing quantum circuits, i.e., inserting SWAP gates.
Transpilers that do not account for gate commutativity in $e^{-i\gamma_k H_C}$ are sub-optimal~\cite{weidenfeller2022scaling}.
Commutation-aware transpiler passes have thus been developed~\cite{Lao2021, Alam2020}.
Predetermined networks of SWAP gates quickly transpile blocks of commuting two-qubit circuits and produce low-depth circuits compared to other methods~\cite{Harrigan2021, weidenfeller2022scaling}.
However, for problems that are not fully connected, such as \textsc{MaxCut} on random-regular-three (RR3) graphs, predetermined swap networks produce even shallower quantum circuits if the initial mapping from the decision variables to the physical qubits is optimized to minimize the number of swap layers~\cite{Matsuo2022}.

In this work, we map the quantum circuits to the best line of qubits on the hardware using alternating layers of SWAP gates~\cite{Harrigan2021, weidenfeller2022scaling}.
The line of qubits is chosen according to the fidelity of the CNOT gates as reported by the backend, see App.~\ref{app:rr3_graphs}.
Furthermore, since RR3 graphs are sparse we reorder the decision variables of the problem to minimize the number of SWAP layers.
This is done by a SAT description of the initial mapping problem~\cite{Matsuo2022}.
Details on the graph generation, transpilation, and SAT mapping are in App.~\ref{app:rr3_graphs}.
We first consider two RR3 graphs with 30 and 40 nodes that can be mapped to the hardware with a total of six and seven swap layers, respectively, once the SAT initial mapping is solved.
The resulting circuits are transpiled to the hardware native gate set $\{X,\sqrt{X}, \text{R}_Z(\theta), {\rm ECR}\}$. 
Here $X$ and $\sqrt{X}$ are the Pauli $X$ gate and its square root.
$\text{R}_Z(\theta)$ is a rotational $Z$ gate with angle $\theta$ and $\rm{ECR}$ is the echoed cross-resonance gate~\cite{Rigetti2010, Sheldon2016}.
The $\rm{ECR}$ gate is equivalent to the standard two-qubit entangling $\rm{CNOT}$ gate up to single-qubit rotations.

\begin{figure}
    \centering
    \includegraphics[width=\columnwidth,clip, trim=5 0 5 0]{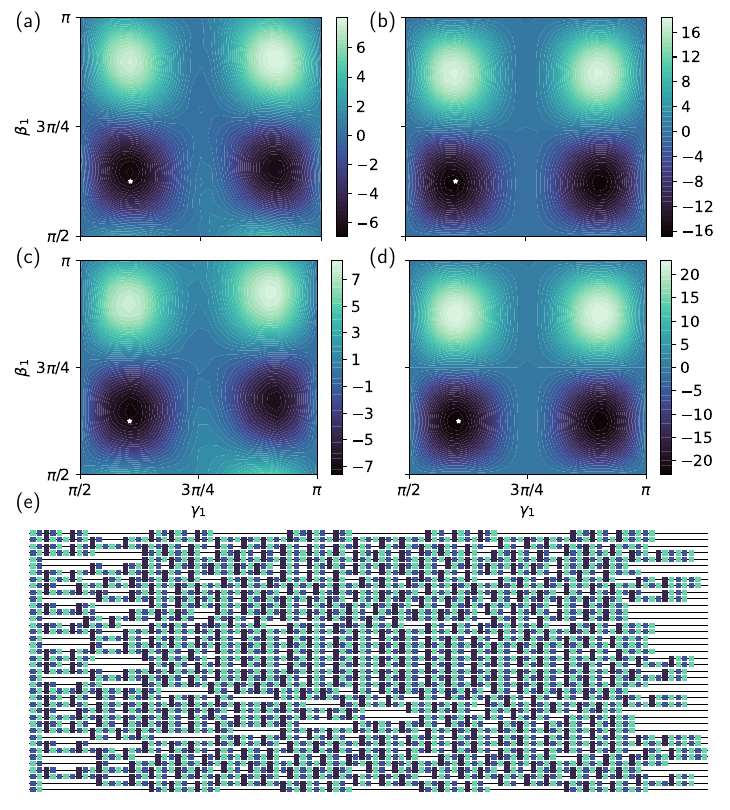}
    \caption{
    (a) and (b) are the depth-one energy landscapes, measured on \emph{ibm\_brisbane} and an ideal simulator, respectively, of a 30 node RR3 graph.
    The circuit (not shown) has 305 ECR gates, 639 $X$ and $\sqrt{X}$ gates and 804 virtual $\text{R}_Z$ gates.
    (c) and (d) are the depth-one energy landscapes, measured on \emph{ibm\_brisbane} and an ideal simulator, respectively, of a 40 node RR3 graph.
    The white stars indicate a minima of the noiseless simulations.
    They reveal a small shift in the corresponding minimum of the hardware-measured data.
    (e) The quantum circuit of the depth-one QAOA of a RR3 40 node graph transpiled to a line of qubits with seven layers of SWAP gates and a SAT-based initial mapping.
    The circuit has 479 ECR gates (dark blue), 1021 $X$ and $\sqrt{X}$ gates (blue), and 1275 virtual $\text{R}_Z$ gates (light green).
    }
    \label{fig:depth_one_qaoa}
\end{figure}

RR3 graphs with 30 and 40 nodes result in large and dense quantum circuits.
For example, a depth-one QAOA creates circuits with 305 and 479 $\rm ECR$ gates for $|V|=30$ and $|V|=40$, respectively, see Fig.~\ref{fig:depth_one_qaoa}(e) which also shows that these circuits leave little space for error suppression methods such as dynamical decoupling~\cite{Viola1999}.
We run the circuits on \emph{ibm\_brisbane} and scan the values $(\gamma_1, \beta_1)$ from $\pi/2$ to $\pi$ in 25 steps to investigate if there is a signal without error mitigation.
We compare the hardware results to an efficient simulation of depth-one QAOA as described in App.~F of Ref.~\cite{egger2021warmstartingquantum}.
The structure of the measured landscape matches the simulations, compare Fig.~\ref{fig:depth_one_qaoa}(a) and (c) to (b) and (d), respectively.
For the 30 and 40 node graphs the contrast, i.e., maximum less minimum, of the hardware-measured landscape is 43.0\% and 33.8\% of the contrast of the simulations, respectively, see the color scales in Fig.~\ref{fig:depth_one_qaoa}.
The location of the hardware and simulation minima are identical in $\gamma$ and shifted in $\beta$ by one grid point, i.e., $65~\mathrm{mrad}$.
Crucially, these results indicate that, despite the large gate count, the quantum computer produces a signal that we can further error mitigate to optimize the parameters of QAOA circuits with $p>1$.

\section{Machine learning assisted error mitigation\label{sec:ml_qem}} 

\begin{figure*}[htbp!]
    \centering
    \includegraphics[width=1.85\columnwidth]{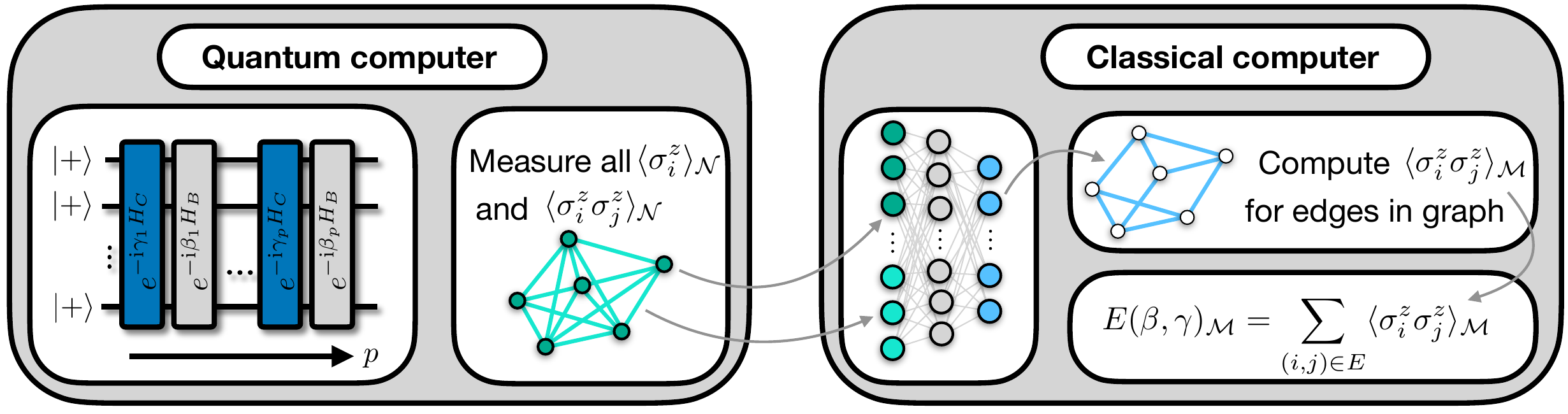}
    \caption{
    Schematic of the error mitigation with a trained FFNN.
    The QAOA circuits of graph $G=(V, E)$ are run and the noisy expectation values $\langle \sigma^z_i\rangle_\mathcal{N}~\forall~i\in V$ and $\langle \sigma^z_i \sigma^z_j\rangle_\mathcal{N}~\forall~i,j\in V$ are computed from the sampled counts.
    These expectation values are the input to a FFNN which outputs the noise mitigated correlators $\langle \sigma^z_i\sigma^z_j\rangle_\mathcal{M}~\forall~(i,j)\in E$.
    The noise mitigated cost function to optimize is $\sum_{(i,j)\in E}\langle \sigma^z_i\sigma^z_j\rangle_\mathcal{M}$.
    }
    \label{fig:NN-cartoon}
\end{figure*}

Inspired by Ref.~\cite{kim2020quantum, strikis2021learning-based, czarnik2021error} we mitigate errors in the energy expectation value with supervised machine learning.
We explore a machine-learning approach based on a neural network to error mitigate QAOA circuits with $p>1$ during the optimization of $\boldsymbol{\gamma}$ and $\boldsymbol{\beta}$.

\subsection{Supervised machine learning}

A supervised machine learning model requires input data $\mathcal{X}= \{X_i\}_{i=1}^{M} $ and target data $\mathcal{Y} = \{Y_i\}_{i=1}^{M}$ to learn the relation between $\mathcal{X}$ and $\mathcal{Y}$ and make predictions on unseen data.
Here, $M$ is the data size.
We build $\mathcal{X}$ from noisy local expectation values and $\mathcal{Y}$ from the corresponding exact, noise-free, expectation values.
The machine learning model learns the relation from noisy data to the noise-free data. 
Our proposed method has three steps.
First, we generate noisy input data $\mathcal{X}$ on a quantum computer.
Second, we simulate the quantum circuits classically to obtain noise-free target data $\mathcal{Y}$.
Finally, we train a machine learning model to learn the mapping from noisy to noise-free data.
The trained model then error mitigates new, i.e., unseen, data. 

\subsection{Feed-forward neural network}

There is a large number of sophisticated supervised machine learning models.
Here, we use a standard fully connected feed-forward neural network (FFNN)~\cite{bishop2007pattern} due to its simplicity and ease of use. 
A FFNN is a series of layers. 
Each layer has multiple neurons that are fully connected to all the neurons in the subsequent layer.
This architecture allows the FFNN to model complex non-linear relationships between the input and output data.
We construct our FFNN with an input layer, a single hidden layer, and an output layer.
Variational algorithms typically minimize the expectation value of a Hamiltonian built from a linear combination of Pauli expectation values $\sum_i\alpha_i P_i$ with $\alpha_i$ a coefficient and $P_i$ a Pauli operator.
To error mitigate a variational algorithm with a FFNN the output layer must yield quantities that can be optimized.
We therefore chose as output layer the correlators that build up the cost function to minimize.
The input is a set of noisy observables measured on the quantum computer.
The FFNN thus maps noisy observables $\langle P_i'\rangle_\mathcal{N}$, measured on hardware, to error mitigated observables $\langle P_i\rangle_\mathcal{M}$. 
The sub-scripts $\mathcal{N}$ and $\mathcal{M}$ indicate noisy and error mitigated observables, respectively.

In the following, we apply the general ideas outlined above to QAOA on a graph $G=(V, E)$ with $|V|=n$ nodes.
We chose an input layer with $n(n+1)/2$ neurons. 
$n$ of these neurons correspond to $n$ noisy local Pauli-Z observables $\langle \sigma_i^z \rangle_{\mathcal{N}}$. 
The other $n(n-1)/2$ neurons correspond to all possible $\langle \sigma_i^z \sigma_j^z \rangle_{\mathcal{N}}$ correlators, where $i,j = 1, 2, ..., n$.
The output layer is made of $|E|$ neurons; one for each correlator $\langle \sigma_i^z\sigma_j^z \rangle_\mathcal{M}$ corresponding to an edge $(i,j)\in E$.
Therefore, a RR3 graph uses a FFNN with $3n/2$ output neurons.
The number of neurons in the hidden layer is the average of the input and output number of neurons.
This construction is illustrated in Fig.~\ref{fig:NN-cartoon}.
A trained FFNN helps us run the QAOA on a quantum computer.
Noisy observables are fed into the FFNN for error mitigation.
The value of the output neurons is summed to produce an error mitigated estimation of the energy expectation value $E(\bm{\beta}, \bm{\gamma})_{\mathcal{M}}=\sum_{i,j \in E} \langle \sigma_i^z \sigma_j^z \rangle_\mathcal{M}$.
This helps optimize $\boldsymbol{\gamma}$ and $\boldsymbol{\beta}$. 

\subsection{Efficient training data generation\label{sec:training_data}}
 
Training the FFNN requires input data $\mathcal{X}$ and target data $\mathcal{Y}$.
We generate $\mathcal{X}$ and $\mathcal{Y}$ by transforming the circuits to error mitigate into classically efficiently simulable circuits.
This can be done in multiple ways.
According to Ref.~\cite{czarnik2021error} it is advantageous to bias the training data towards the state of interest.
The QAOA seeks the ground state of $H_C$, typically a classical product state.
It applies the unitaries $e^{-i \beta_k H_B}$ and $e^{-i \gamma_k H_C}$ to drive the initial equal superposition $\ket{+}^{\otimes n}$ towards the ground state of $H_C$.
To generate the training data we could restrict the angles $\beta_k$ and $\gamma_k$ to reduce $e^{-i \beta_k H_B}$ and $e^{-i \gamma_k H_C}$ to Clifford circuits.
This would however result in a small training data set and may not be possible if the edges in $E$ have non-integer weights.
Alternatively, we could randomly replace each $\text{R}_Z$ rotation in the transpiled circuit of $e^{-i \gamma_k H_C}$ by a Clifford gate such as $I$, $S$, and $Z$.
However, this alters the graph $G$ by giving the edges in $E$ random weights.
This may be undesirable as the structure of the QUBO of interest is changed.

\begin{figure}[h]
    \centering
    \includegraphics[width=\columnwidth,clip,trim=10 0 0 0]{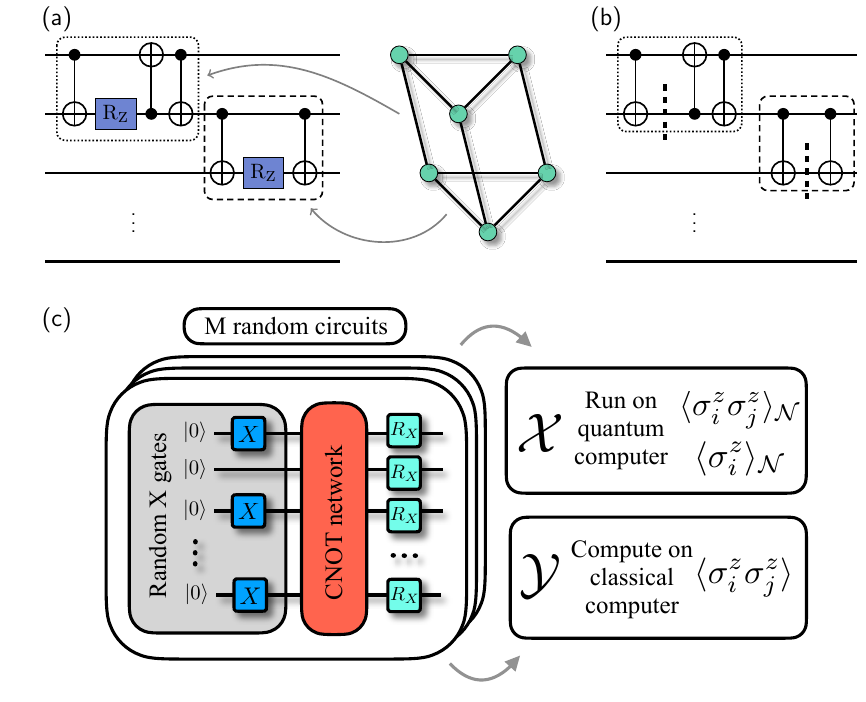}
    \caption{
        (a) Part of a QAOA cost operator.
        The dotted and dashed gates correspond to an $\text{R}_{ZZ}$ gate with and without a SWAP gate transpiled to CNOT gates.
        (b) The $\text{R}_{Z}$ gates are replaced by barriers to prevent CNOT gate cancellation.
        This preserves the noise structure of the circuit.
        (c) Training data generation.
        $M$ random input cuts are created by randomly applying $X$ gates to the qubits.
        These states are propagated through $p$ alternating networks of CNOT gates, corresponding to simplified $e^{-i\gamma H_C}$ operators, and mixer layers $e^{-i \beta H_B}$. 
        These circuits are run on hardware to create the input data $\mathcal{X}$ and efficiently simulated classically to generate the ideal output data $\mathcal{Y}$.
        }
    \label{fig:remove-rz}
\end{figure}

These considerations motivate us to train the FFNN on data obtained by sampling over random product states that have undergone a noise process qualitatively similar to the QAOA without altering $G$.
First, we change the initial state from an equal superposition to a random partition of $V$ by randomly applying $X$ gates to the qubits.
This initial state is followed by circuit instructions that generate noise similar to the noise in the QAOA.
The cost operator (up to SWAP gates which we omit in the following for simplicity) is
\begin{equation}
    e^{-i \gamma_k H_C} = \prod_{(i, j) \in E} e^{-i \gamma_k \sigma_i^z \sigma_i^z} =  \prod_{(i, j) \in E} \text{CX}_{i, j} \text{R}^j_Z(2\gamma_k) \text{CX}_{i, j}.
\end{equation}
Where, $\text{CX}_{i, j}$ is a CNOT gate between qubits $i$ and $j$ and $\text{R}^i_Z$ is a rotation around the Z axis of qubit $i$.
By setting $\gamma_k=0$ the operator $e^{-i\gamma_k H_C}$ reduces to the identity (up to SWAP gates) and the QAOA circuit produces product states that we efficiently simulate classically.
To retain the noise characteristics, we replace the $\text{R}^i_Z$ gates with barriers to prevent the transpiler from removing the CNOT gates, see Fig.~\ref{fig:remove-rz}. 
Since $\text{R}^i_Z$ is implemented by virtual phase changes~\cite{mckay2017efficient}, the duration and magnitude of all pulses played on the hardware are unchanged. 
This preserves the effect of $T_1$, $T_2$, cross-talk, and other forms of errors.
In detail, we generate training data with a set of $M$ random states
\begin{equation}
\ket{\bm{\beta}}=\prod_{i=1}^p  e^{-i \beta_i H_B} \!\! \prod_{(k,l) \in E}\!\!\text{CX}_{k,l}^2  \bigg[p_j X\ket{0}_j+(1-p_j)\ket{0}_j \bigg]^{\otimes n},
\end{equation}
that are used to measure the observables for the input data $\mathcal{X}$. 
Here, each $\beta_i$ is a uniform random variable in $[0,2\pi]$ and $p_j$ is a Bernoulli random binary variable that applies an $X$ gate on qubit $j$ if successful.
We chose a $1/2$ probability of success for $p_j$.
To compute the target data we use $\text{CX}_{k,l}^2=\mathbb{I}$, the resulting state is a trivial product state for which it is straightforward to efficiently compute the exact expectation values required for the target data~$\mathcal{Y}$.

\section{Machine learning error mitigated QAOA\label{sec:ml_qaoa}}

We now apply the FFNN error mitigation discussed in Sec.~\ref{sec:ml_qem} and the QAOA execution methods discussed in Sec.~\ref{sec:qaoa} to run depth-two QAOA.
We first exemplify the error mitigation in a small ten qubit simulation and then turn to larger RR3 graphs with ten, twenty, thirty, and forty nodes executed on hardware.

\subsection{Simulations}

We build a noise model with short-lived qubits.
Their $T_1$ and $T_2$ times are sampled from a Gaussian distribution with $10~\mu {\rm s}$ mean and $10~{\rm ns}$ standard deviation.
Based on these durations, a thermal relaxation noise channel is applied to the CNOT gates lasting $\tau_\text{\tiny CNOT}=300~{\rm ns}$.
This is a strong noise model for the 102 CNOT gates in the QAOA circuit as understood, e.g., by $e^{-\tau_\text{\tiny CNOT}/T_1}$ which gives 97\% as proxy for the gate fidelity. 
The other circuit instructions are noiseless.

\begin{figure}
    \centering
    \includegraphics[width=\columnwidth, clip, trim=8 5 8 25]{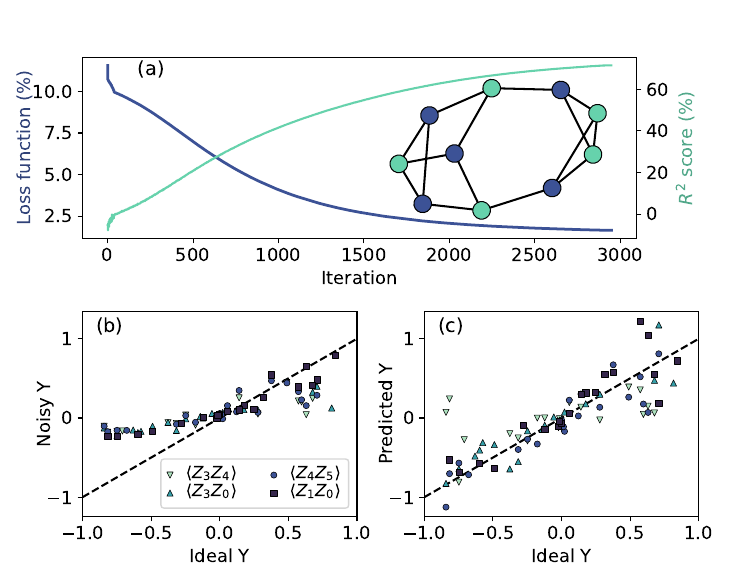}
    \caption{
    Training of a FFNN on simulated data.
    (a) Loss function (dark blue) and $R^2$ score (light green) as a function of the training iteration.
    The inset is the ten-node RR3 graph with node color indicating the \textsc{MaxCut}.
    (b) Correlators corresponding to edges in the inset graph in (a) before error mitigation versus their ideal value.
    If the device was noiseless, all the correlators would lie on the dashed line.
    (c) The correlators of (b) but after error mitigation by the FFNN.
    If the error mitigation were perfect, all the correlators would lie on the dashed line.
    }
    \label{fig:nn_sim}
\end{figure}

We sample 300 random cuts to create the training data following Sec.~\ref{sec:training_data}.
We train the FFNN with 90\% of this data and the other 10\% serves as validation data~\footnote{The FFNN is implemented with the \texttt{MLPRegressor} from sklearn.}.
In this example, the FFNN achieves a mean squared error (MSE) of 3.3\% on the training data and an $R^2$ score of 71.8\% on the validation data, see Fig.~\ref{fig:nn_sim}(a).
The FFNN thus captures 71.8\% of the variation in the validation data.
Furthermore, we generate an additional 20 test data points. 
The corresponding non-error mitigated correlators are damped towards an expectation value of zero, see Fig.~\ref{fig:nn_sim}(b).
The MSE between these $20\times |E|=300$ non-error-mitigated and ideal $\langle \sigma^z_i\sigma^z_j \rangle$ correlators is 11\%.
This number drops to 7\% after the correlators are error mitigated with the FFNN.
Furthermore, we observe that the error-mitigated correlators better follow the trend set by their ideal values, see Fig.~\ref{fig:nn_sim}(c). 

In a separate simulation, we increase the strength of the noise by lengthening the CNOT gates.
At a duration of $400~{\rm ns}$ the FFNN cannot learn an error mitigation since the noise is too strong.
We observe that the squared error does not reach low values and the predicted correlators are close to zero (data not shown).

\begin{figure}
    \centering
    \includegraphics[width=\columnwidth]{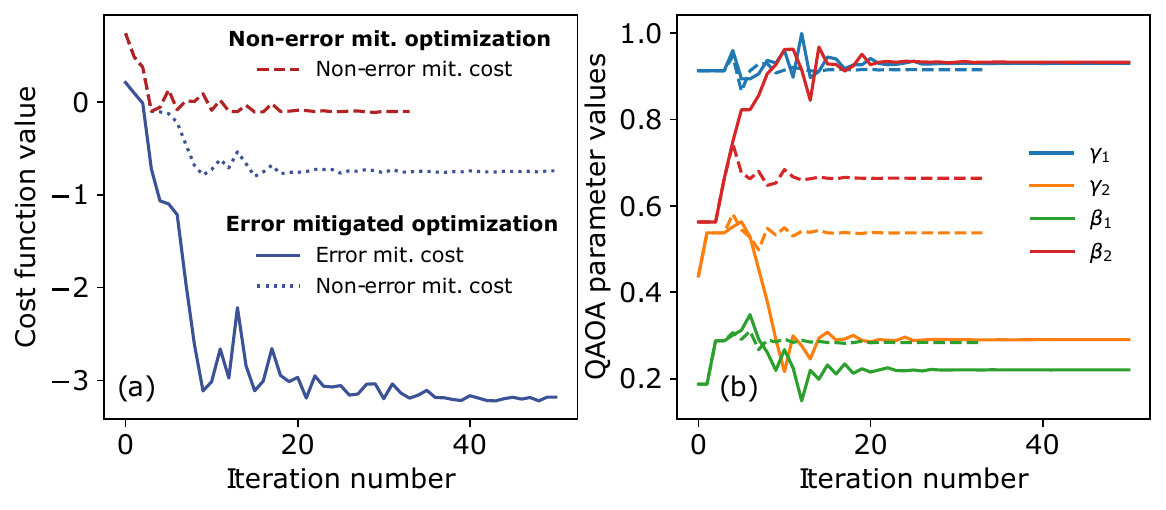}
    \caption{
    Optimization of a noisy QAOA.
    (a) The solid and dotted lines show the error mitigated cost function and the non-error mitigated cost function, respectively during the optimization of the former.
    The dashed line shows the non-error mitigated cost function as it is being optimized.
    (b) QAOA parameters for the error mitigated optimization (solid lines) and the non-error mitigated optimization (dashed lines).
    }
    \label{fig:sim_qaoa}
\end{figure}

We now optimize a $p=2$ QAOA with $300~{\rm ns}$ CNOT gates twice; once by optimizing the error mitigated cost function $E_\mathcal{M}=\sum_{(i,j)\in E}\langle\sigma_j^z\sigma_j^z\rangle_\mathcal{M}$, blue data in ~Fig.~\ref{fig:sim_qaoa}(a), and once by optimizing the non-error mitigated cost function $E_\mathcal{N}=\sum_{(i,j)\in E}\langle\sigma_j^z\sigma_j^z\rangle_\mathcal{N}$, red data in ~Fig.~\ref{fig:sim_qaoa}(a).
We use COBYLA with $\boldsymbol{\theta}=(\gamma_1, \gamma_2, \beta_1, \beta_2)$ initialized from a Trotterized Quantum Annealing schedule~\cite{Sack2021}.
Each circuit is run with 4096 shots.

\begin{figure}
    \centering
    \includegraphics[width=\columnwidth, clip, trim=4 5 5 5]{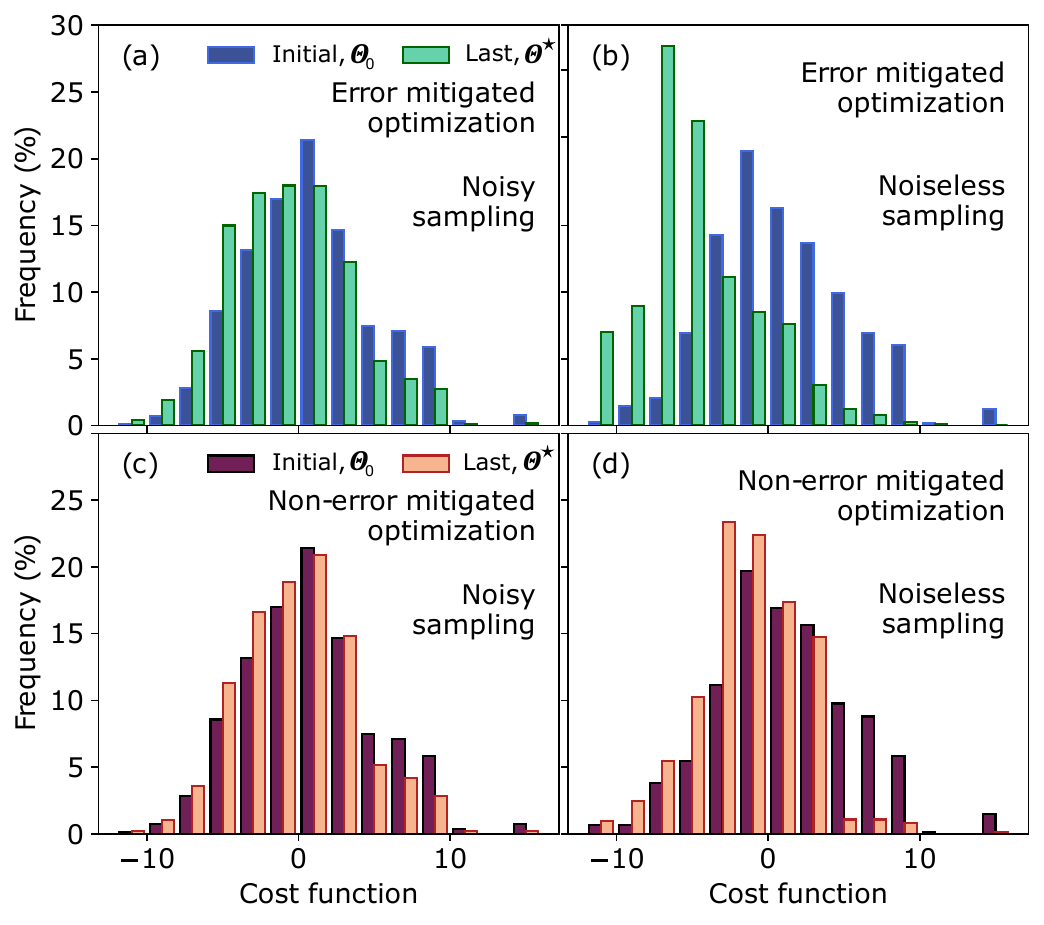}
    \caption{
        Distribution of the cost function of each individual sampled bitstring for the optimization in Fig.~\ref{fig:sim_qaoa}.
        (a) and (b) correspond to the optimization with the error mitigated cost function (solid lines in Fig.~\ref{fig:sim_qaoa}).
        (c) and (d) correspond to the optimization with the non-error mitigated cost function (dashed lines in Fig.~\ref{fig:sim_qaoa}).
    }
    \label{fig:sim_sampling}
\end{figure}

The error mitigated energy reaches lower values than the non-error mitigated energy, see Fig.~\ref{fig:sim_qaoa}(a).
Both optimizations converge within 40 iterations, see Fig.~\ref{fig:sim_qaoa}(b).
Furthermore, when the error mitigated cost function is optimized, the corresponding non-error mitigated cost function, dotted blue curve in Fig.~\ref{fig:sim_qaoa}(a), reaches lower values than a direct optimization of the non-error mitigated cost function, red dashed curve in Fig.~\ref{fig:sim_qaoa}(a).
This shows that with the error mitigation on, the optimizer finds better values of $\boldsymbol{\theta}$.
To further illustrate this, we compute the energy distribution of sampled bitstrings.
We compare the distribution of the cost function of each sampled bitstring of the initial and last values of $\boldsymbol{\theta}$ labeled $\boldsymbol{\theta}_0$ and $\boldsymbol{\theta}^\star$ respectively.
The sampling is done with the noisy simulator, Fig.~\ref{fig:sim_sampling}(a) and (c), and a noiseless simulator, Fig.~\ref{fig:sim_sampling}(b) and (d).
Sampling from a noisy $\ket{\boldsymbol{\theta}^\star}$ produces a distribution that is near identical to the one obtained by sampling from a noisy $\ket{\boldsymbol{\theta}_0}$, see Fig.~\ref{fig:sim_sampling}(a) and (c).
However, sampling from a noiseless $\ket{\boldsymbol{\theta}^\star}$ produces a better distribution than sampling from a noiseless $\ket{\boldsymbol{\theta}_0}$, see Fig.~\ref{fig:sim_sampling}(a) and (b).
This suggests that the error mitigation helps find better values of $\boldsymbol{\theta}$ despite the fact that we cannot see this by \emph{sampling} bitstrings from noisy QAOA states.

Finally, we repeat these simulations 20 times.
In each simulation we train a FFNN and optimize $\boldsymbol{\theta}$.
This produces different optimization results due to the randomness of the noise.
The optimization is carried out twice, once on $E_\mathcal{M}$ and once on $E_\mathcal{N}$.
After the optimization we sample 4096 bitstrings from a noiseless simulation of the QAOA circuit with the optimized parameters $\boldsymbol{\theta}^\star$.
We compute the energy distribution of these bitstrings and report the expectation value.
This expectation value is $-3.78\pm2.12$ and $-2.63\pm2.01$ when $E_\mathcal{M}$ and $E_\mathcal{N}$ is optimized, respectively.
These results indicate that error mitigation tends to help the classical optimizer find better QAOA parameter values.
However, we observe in seven out of twenty simulations that an optimization of the noisy QAOA cost function $E_\mathcal{N}$ produces better parameters, as measured by a noiseless sampling of bitstrings, than an optimization of the error mitigated cost function $E_\mathcal{M}$.
In all simulations, except one, the error mitigated cost function $E_\mathcal{M}$ has a lower energy than the non-error mitigated cost function $E_\mathcal{N}$.

\subsection{Hardware}

\begin{figure*}
    \centering
    \includegraphics[width=\textwidth, clip,trim=1500 150 1100 150]{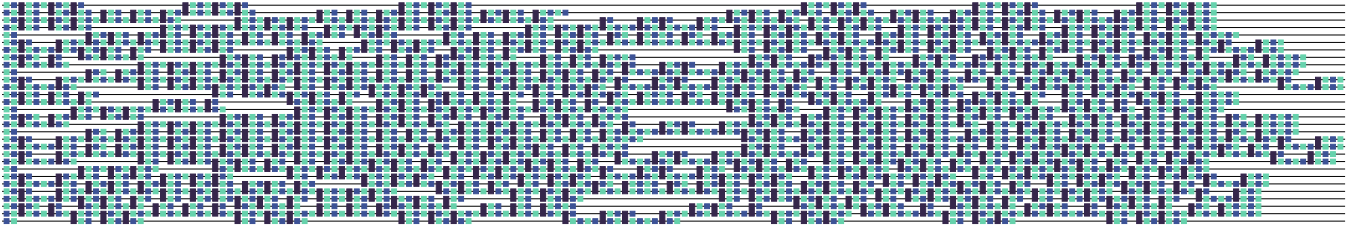}
    \caption{
    Depth-two QAOA circuit of a 30 node RR3 containing 610 ECR gates (dark blue), 1297 single-qubit $X$ and $\sqrt{X}$ gates (blue), and 1577 virtual $\text{R}_Z$ gates (teal).
    }
    \label{fig:depth2}
\end{figure*}

We now optimize the parameters $\gamma_1$, $\gamma_2$, $\beta_1$, and $\beta_2$ with COBYLA of a depth-two QAOA for RR3 graphs on superconducting qubit hardware.
As cost function we minimize the energy $E_\mathcal{M}=\sum_{(i,j)\in E}\langle\sigma_i^z\sigma_j^z\rangle_\mathcal{M}$ computed with error mitigated correlators produced by FFNNs.
Before each run, the FFNN is trained, as described in Sec.~\ref{sec:ml_qem}, with 3000 training points evaluated with 1024 shots each.

\begin{figure*}
    \centering
    \includegraphics[width=\textwidth]{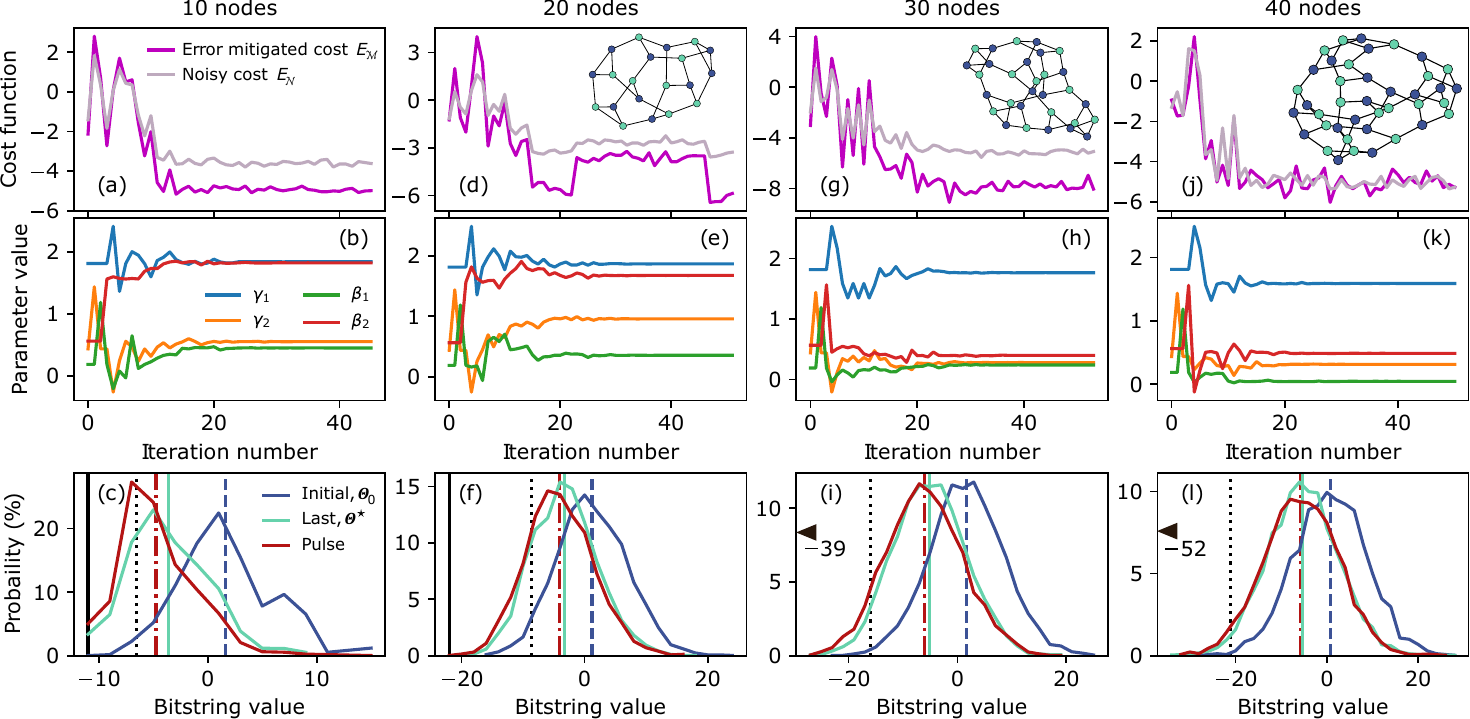}
    \caption{
        Depth-two QAOA data acquired on \emph{ibm\_brisbane}.
        The graphs are shown as insets with node colors indicating the \textsc{MaxCut} found with CPLEX~\cite{cplex}. 
        (a), (d), (g), and (j) show the error mitigated cost function $E_\mathcal{M}$ obtained from FFNNs at each iteration of COBYLA.
        The non-mitigated cost function $E_\mathcal{N}$, is also shown as $E_\mathcal{M}$ is optimized. 
        $E_\mathcal{N}$ and $E_\mathcal{M}$ thus represent the same quantities as the blue dotted and blue solid lines in Fig.~\ref{fig:sim_qaoa}(a), respectively.
        (b), (e), (h), and (k) show the QAOA parameters during the optimization.
        (c), (f), (i), and (l) show the distribution of the energy of the sampled bit strings.
        The dark blue and light teal lines correspond to bitstrings sampled from the QAOA circuits with the initial and last parameters $\boldsymbol{\theta}_0$ and $\boldsymbol{\theta}^\star$, respectively.
        The red lines correspond to sampling from a pulse-efficient circuit.
        The solid black lines in (c) and (f) indicate the energy of the \textsc{MaxCut}.
        In (i) and (l) this energy lies outside of the x-axis range.
        We indicate it as a triangle with the energy as a number. 
        The black dotted vertical lines indicate the noiseless expectation value obtained with $\boldsymbol{\theta}^\star$.
        }
    \label{fig:study}
\end{figure*}

For RR3 graphs with 30 nodes the quantum circuits have a total of 610 ECR gates, 1297 $X$ and $\sqrt{X}$ gates, and 1577 $R_Z$ gates, see Fig.~\ref{fig:depth2}.
Graphs with 40 nodes have a total of 958 ECR gates.
While these circuits are extremely deep and wide they still contain a significant signal.
When running the QAOA optimization on hardware we observe a minimization of $E_\mathcal{M}$ for all graphs, see dark purple curves in Fig.~\ref{fig:study}(a), (d), (g), and (j).
The non-error mitigate cost function $E_\mathcal{N}$ (light purple curves) also decreases.
In all cases the optimization of $\gamma_i$ and $\beta_i$ converges in about 20 to 40 iterations of COBYLA, see Fig.~\ref{fig:study}(b), (e), (h), and (k).
We compare the distribution of the sampled bitstrings obtained from QAOA circuits with the initial points $\boldsymbol{\theta}_0$ to the distribution obtained with the optimized $\boldsymbol{\theta}^\star$.
We see an improvement in the distribution, i.e., a bias towards lower values, for all RR3 graphs, compare the dark blue and light teal curves in Fig.~\ref{fig:study}(c), (f), (i) and (l).
This is consistent with the interpretation that there is a meaningful signal in the corresponding circuits.
We report the mean $\mu$ of each distribution (vertical lines in Fig.~\ref{fig:study}) as an approximation ratio $\alpha(\mu)=(\mu - \langle H_C\rangle_\text{max})/(\langle H_C\rangle_\text{min}-\langle H_C\rangle_\text{max})$ contained in $[0, 1]$.
The optimized parameters produce an $\alpha(\mu)$ of 71.6\%, 64.0\%, 59.6\%, and 58.3\% for the 10, 20, 30, and 40 node graphs.

We optimize the parameters $\boldsymbol{\theta}$ with ECR-based circuits since all parameters are in virtual $\text{R}_Z$ gates.
This preserves the amplitude and duration of all pulses in the schedule thus facilitating noise mitigation.
Pulse-efficient transpilation moves the parameters from the $\text{R}_Z$ gates into the cross-resonance pulses~\cite{Earnest2021, Stenger2021}.
This shortens the pulse schedule but changes its noise properties.
For example, a pulse-efficient transpilation of an $\text{R}_{ZZ}(\theta)$ - SWAP pair, as shown in Fig.~\ref{fig:remove-rz}(a), reduces their duration by up to 20\%, depending on $\theta$.
The shorter schedules produce better bitstrings than the fixed-duration schedules with parameters in $\text{R}_Z$ gates. 
We run the pulse-efficient circuits for the last points $\boldsymbol{\theta}^\star$.
This results in an improved $\alpha(\mu)$ of 76.0\%, 65.5\%, 60.8\%, and 58.6\% over the same circuit without pulse-efficient transpilation for the 10, 20, 30, and 40 node graphs, respectively, compare the dash-dotted red line to the solid teal line in Fig.~\ref{fig:study}(c), (f), (i) and (l).

The best bit strings sampled from the distributions shown in Fig.~\ref{fig:study}(c), (f), (i) and (l) have a cut value of 13 (\textsc{Maxcut}), 26 (\textsc{Maxcut}), 35 (0.833), and 44 (0.786) for the 10, 20, 30, and 40 node graphs, respectively.
The numbers in parenthesis indicate the approximation ratio.
These bitstrings were observed a total of 1619, 124, 474, and 988 times, respectively, out of the $3\times 4096$ shots in the distributions.

To distinguish the impact of hardware noise on the bitstring distribution from limitations of the depth-two QAOA Ansatz, we compute the noiseless expectation value of the cost Hamiltonian $H_C=\sum_{(i, j)\in E}\sigma_i^z\sigma_j^z$.
We evaluate $H_C$ at the last point $\boldsymbol{\theta}^\star$ obtained from the noisy hardware optimization.
This computation is made fast, even for a 40 node graph, with quantum circuits based on the light-cone of each correlator $\sigma_i^z\sigma_j^z$.
This method is detailed in App.~\ref{app:light_cone}.
The noiseless expectation value is indicated as a dotted line in Fig.~\ref{fig:study}(c), (f), (i) and (l).
The corresponding approximation ratios $\alpha(\mu)$ are 82.8\%, 74.2\%, 72.6\%, and 72.4\% for the 10, 20, 30, and 40 node graphs, respectively.
These values show the potential improvement in the bitstring value distribution if hardware noise could be reduced.


\section{Discussion and Conclusion\label{sec:conclusion}}

Many machine learning tools can error mitigate an expectation value.
The first contribution of this work is a user firendly FFNN-based error mitigation strategywith a problem-inspired methodology to generate the training data.
Here, the FFNN is trained once before the variational optimization.
It tries to match training data acquired at randomly sampled values of $\beta$ for which all values of $\gamma$ are set to zero with $R_Z$ gates replaced by barriers.
This preserves the circuit structure and makes it easy to simulate classically.
We observe that the FFNN performs better on validation data than a linear regression as the circuit size is increased, see App.~\ref{sec:ml_comp}.
The data from the optimization with error mitigation show that the FFNN reduces the effect of noise on the cost function at unseen values of $\beta$ and~$\gamma$.
Other data generation approaches are possible and could be investigated in future work which may also explore other machine learning tools such as random forests as done in Ref.~\cite{Liao2023}.

\begin{figure}
    \centering
    \includegraphics[width=\columnwidth]{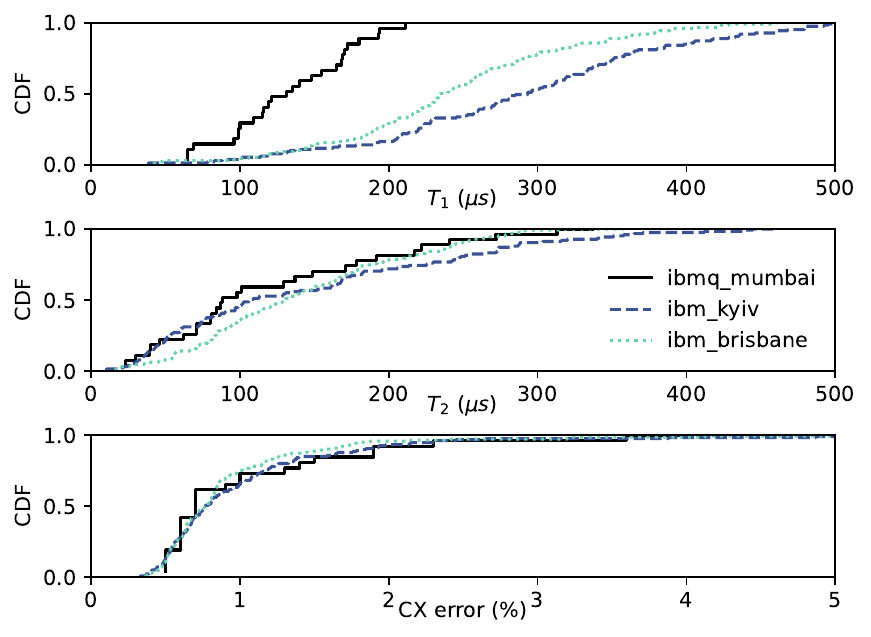}
    \caption{
    Properties of the Eagle quantum processors used in this paper, \emph{ibm\_brisbane} and \emph{ibm\_kyiv}, compared to the Falcon processor used in Ref.~\cite{weidenfeller2022scaling}, i.e., \emph{ibmq\_mumbai}.
    The data are presented as cumulative distribution functions (CDF).
    }
    \label{fig:hardware}
\end{figure}

Our second contribution is to implement non-planar RR3 graphs on hardware by leveraging the SAT mapping of Matsuo \emph{et al.}~\cite{Matsuo2022} and swap networks~\cite{Harrigan2021, weidenfeller2022scaling}.
We observe a meaningful signal for a depth-two QAOA with up to 40 nodes.
The swap networks with 2, 4, 6, and 7 layers that we implement on 10, 20, 30, and 40 qubits can generate graph densities of up to 40\%, 30\%, 27\%, and 22\%, respectively.
The corresponding circuits have impressive gate counts.
We attribute hardware improvements of the Eagle quantum processors, shown in Fig.~\ref{fig:hardware}, to this success.
For example, the $T_1$ times of \emph{ibm\_brisbane} are more than twice as large as those of \emph{ibmq\_mumbai}, see Fig.~\ref{fig:hardware}(a), which was used in the 27 qubit experiment in Ref.~\cite{weidenfeller2022scaling}.
The cumulative distribution function of the gate error of \emph{ibm\_brisbane} and \emph{ibmq\_mumbai} is approximatively the same, see Fig.~\ref{fig:hardware}(c).
However, \emph{ibm\_brisbane} has 127 qubits while \emph{ibmq\_mumbai} only has 27 qubits.
this allows us to run the 40 qubit RR3 graph on the \emph{best} line of qubits.
For example, the product of the fidelity of the 39 gates on the best line with 40 qubits on \emph{ibm\_brisbane} is 76.5\%, see App.~\ref{app:rr3_graphs}.
By contrast, the product of all 28 two-qubit gate fidelities on \emph{ibmq\_mumbai} only reaches 72.8\% despite there being fewer gates.

Circuits with 40 nodes can be simulated classically.
In particular, computing $\langle H_C\rangle$ of a depth-$p$ QAOA for a RR3 graph produces an effective light cone with at most $\sum_{k=0}^p 2^{k+1}$ qubits in the circuits, i.e., 14 in our $p=2$ case.
This allows us to confirm that optimizing the error mitigated correlators produces good variational parameters.
As quantum computers increase in size and quality, such a classical verification will no longer be possible.
Crucially, our hardware demonstration is much larger than many current experiments which typically employ up to 20 qubits~\cite{Ichikawa2023}.
Furthermore, reproducing even depth-one QAOA samples is classically intractable~\cite{Farhi2019}.
Our work is thus a step towards implementing QAOA on hardware that cannot be classically simulated.
Future work must focus on implementing deeper circuits on hardware with more connectivity.
Our work also serves as a benchmark to track quantum hardware progress, as done, e.g., with complete graphs~\cite{Santra2022}.
We anticipate that hardware improvements, e.g., increasing $T_1$ times, and novel architectures, based on, e.g., tunable couplers~\cite{Ganzhorn2020, Sung2021}, will enable larger simulations.

The depth-one QAOA results, shown in Sec.~\ref{sec:qaoa}, exhibit the parameter concentration already observed in the literature~\cite{Brandao2018, Akshay2021, Galda2021, Streif2020, Shaydulin2022}.
This may be important to quickly generate good yet sub-optimal solutions to combinatorial optimization problems without having to optimize the variational parameters for each problem instance~\cite{weidenfeller2022scaling}.

The objective of QAOA is to \emph{sample} good or even optimal solutions from a quantum state $\ket{\boldsymbol{\theta}^\star}$ that minimizes the energy $H_C$, i.e., to find $x_{\text{opt}} = {\rm argmin}_{x} H_C(x)$.
The state is obtained by minimizing the \emph{expectation} value of $H_C$.
The error mitigation method we present helps find good parameters $\boldsymbol{\gamma}$ and $\boldsymbol{\beta}$.
However, quantum approximate optimization needs tools that error mitigate \emph{samples}.
Noiseless simulations of $\langle H_C\rangle$ computed with the optimal parameters show that a significant gain is obtainable if samples could be error mitigated.
Recently, Barron \emph{et al.} show that a noise dependent sampling overhead produces good solution samples~\cite{Barron2023}.
This makes sense in the context of optimization as long as the total samples drawn is at least less than the $2^n$ cost of a brute-force search and ideally less than the sampling complexity of the best classical benchmark.
Hardware improvements in, e.g., the fidelity of a layer of gates~\cite{Mckay2023}, and noise-aware transpilers~\cite{Murali2019, Wagner2024}, will reduce this sampling overhead.
Furthermore, proper QAOA benchmarks must compare to state-of-the-art solvers~\cite{Rehfeldt2023}, such as Gurobi and CPLEX, and randomized rounding algorithms~\cite{Goemans1995}.
For example, for RR3 graphs there is an approximation algorithm that achieves an approximation ratio of 0.9326~\cite{Halperin2004}.
Such benchmarking is an important task in itself which our methods enable on hardware.

For variational algorithms, like the variational eigensolver applied in a chemistry setting~\cite{Peruzzo2014, Moll2018}, error mitigating expectation values are often sufficient, e.g., to compute the energy spectrum of molecules~\cite{Ganzhorn2019, Ollitrault2020}.
The FFNN-based error mitigation method we present is directly transferable to such settings which increases its applicability.
Furthermore, the transpiler methodology we leverage works with any non-hardware-native block of commuting two-qubit gates.
It thus applies to circuits other than QAOA such as graph states~\cite{Mooney2021}, and algorithms that implement $e^{-iH_Ct}$ including Ising simulations~\cite{Vazquez2022}.

\section*{Acknowledgments}

IBM, the IBM logo, and ibm.com are trademarks of International Business Machines Corp., registered in many jurisdictions worldwide. Other product and service names might be trademarks of IBM or other companies. The current list of IBM trademarks is available at \url{https://www.ibm.com/legal/copytrade}. S.H.S. acknowledges support from the IBM Ph.D. fellowship 2022 in quantum computing. The authors also thank M. Serbyn, R. Kueng, R. A. Medina, and S. Woerner for fruitful discussions. 
The code for these results is available at \url{https://github.com/eggerdj/large_scale_qaoa}.


\appendix

\section{RR3 graph transpilation\label{app:rr3_graphs}}

Random regular graphs are sparse.
When their corresponding QAOA circuit is transpiled to the hardware with predetermined swap layers certain edges may require a large number of swap layers.
By wisely choosing the initial mapping between decision variables and physical qubits we reduce the number of swap layers needed.
A SAT based approach to this ``initial mapping'' problem is proposed by Matsuo \emph{et al.}~\cite{Matsuo2022}.
Here, the initial mapping problem is formulated as a SAT problem that is satisfiable if $e^{-i\gamma H_C}$ can be routed to hardware with $\ell$ swap layers. 
A binary search over $\ell$ finds the initial mapping that minimizes the number of layers.
We label the minimum number of swap layers by $\ell^*$.

\subsection{Graph generation}

We generate 100 RR3 graphs with $n$ nodes for each $n\in \{10,20, 30, 40\}$. 
Each graph is mapped to a line of $n$ qubits with the SAT approach.
The distribution of the number of swap layers at the different sizes $n$ is shown in Tab.~\ref{tab:sat_res}.
With the SAT mapping, there are graph instances with 10, 20, 30, and 40 nodes that can be implemented with 2, 4, 6, and 7 swap layers, respectively.
This is a large reduction compared to a trivial mapping which typically requires $n-2$ swap layers~\cite{Matsuo2022}.
The experiments in the main text are done on graphs that require the smallest number of swap layers.

\begin{table}[htbp!]
    \centering
    \begin{tabular}{l | r r r r r r r r r} \hline\hline
                Number & \multicolumn{9}{c}{Number of swap layers $\ell^*$} \\ 
        of nodes &  2 & 3 & 4 & 5 & 6 & 7 & 8 & 9 & 10 \\ \hline
        10 & $\quad$26 & $\quad$40 & 34  \\
        20 &    &    & $\quad$13 & $\quad$55 & $\quad$32 \\
        30 &    &    &    &    & 18 & $\quad$64 & $\quad$18 \\
        40 &    &    &    &    &    &  3 & 17 & $\quad$67 & $\quad$13 \\ \hline\hline
    \end{tabular}
    \caption{Number of RR3 graphs that required $\ell^*$ swap layers after an initial mapping found through solving SAT problems.
    At each graph size with $n$ nodes, 100 graph instances were generated.
    For instance, out of 100 random instances of RR3 graphs with 40 nodes three could be mapped to a line of qubits with 7 swap layers.
    }
    \label{tab:sat_res}
\end{table}

\subsection{Qubit selection}

\emph{ibm\_brisbane} has 127 qubits, i.e., 87 more than the largest graph we study.
We, therefore, select the best line of qubits to execute the quantum circuits on.
For each pair of nodes $i$ and $j$ in the backend's coupling map, we enumerate all paths of length $|V|$ connecting them.
Next, we compute the path fidelity for each path $p_k$ as $\prod_{(i,j)\in p_k}(1-E_{\text{ECR}, i,j})$ and select the best one.
Here, $E_{\text{ECR}, i,j}$ is the error of the ECR gate between qubits $i$ and $j$.
On \emph{ibm\_brisbane} there are 1336, 15814, 125918, and 754462 lines of 10, 20, 30, and 40 qubits, respectively.
The best measured respective path fidelities are 95.9\%, 89.5\%, 82.8\%, and 76.5\%.

\begin{figure}
    \centering
    \includegraphics[width=\columnwidth, clip, trim=0 30 0 30]{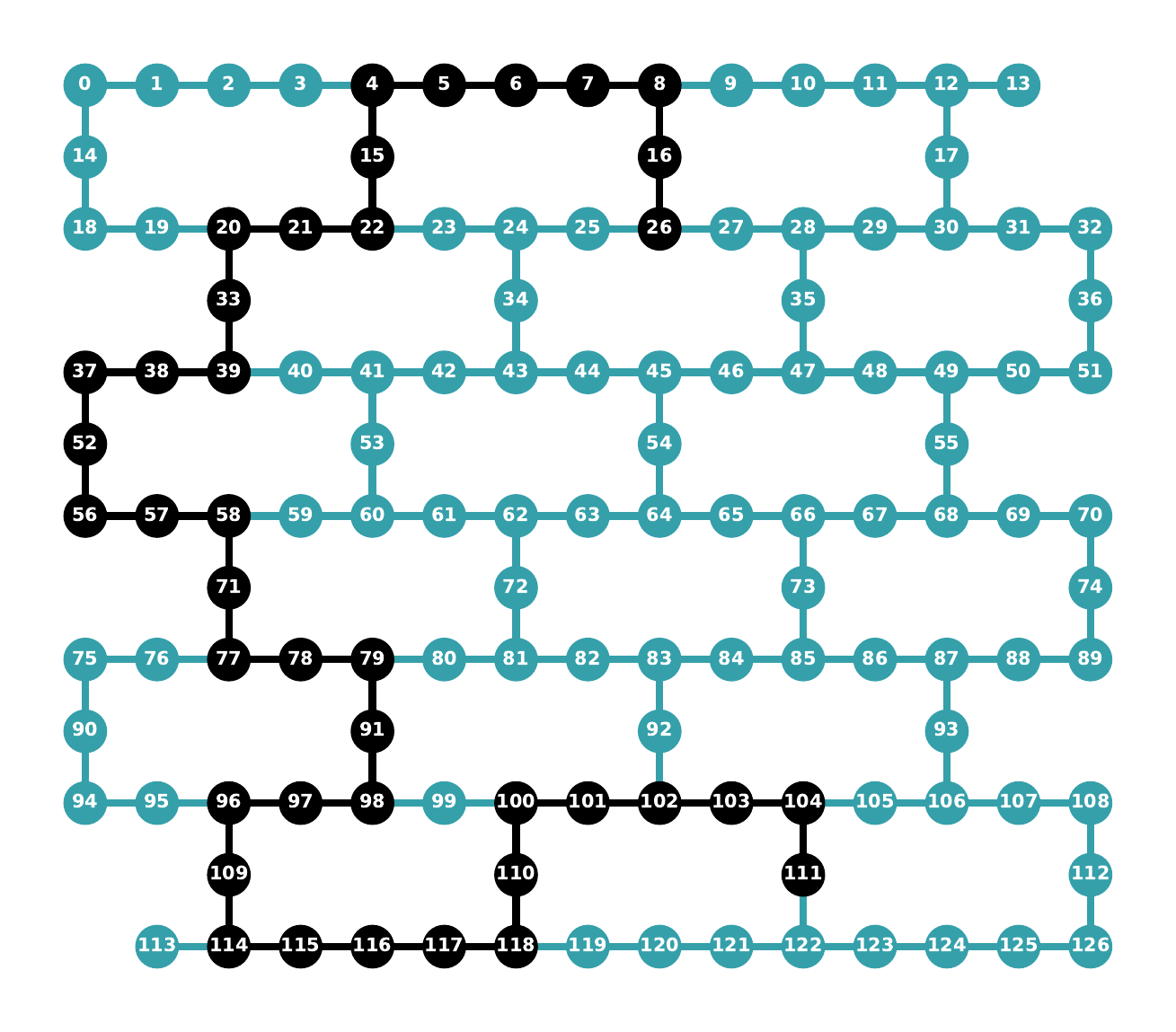}
    \caption{
        Coupling map of \emph{ibm\_brisbane} showing the qubit connectivity.
        The black qubits form the 40 qubits line with the best path fidelity as measured by the quality of the ECR gates.
    }
    \label{fig:brisbane_path}
\end{figure}

\section{Light-cone QAOA\label{app:light_cone}}

\begin{figure}
    \centering
    \includegraphics[width=\columnwidth, clip, trim=10 0 0 0]{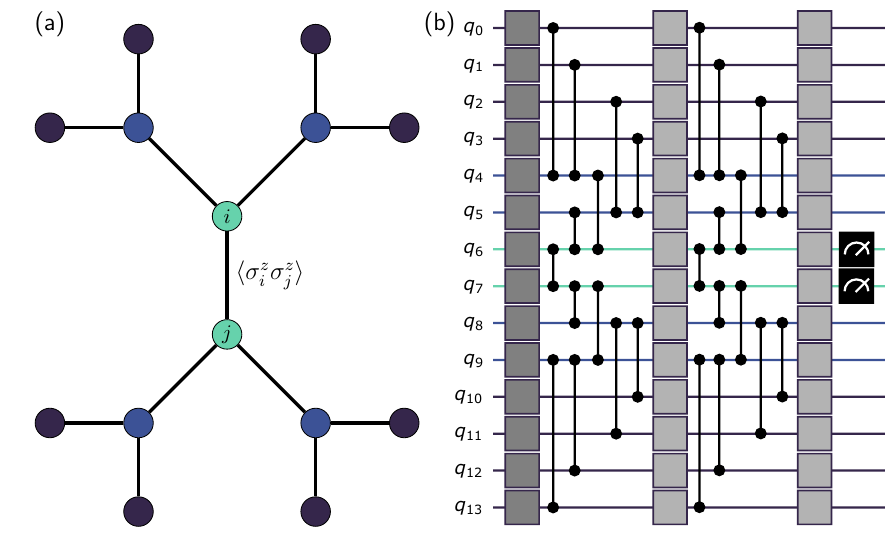}
    \caption{
    (a) Sub-graph of a RR3 graph with all nodes up to a distance two from the nodes $i$ and $j$ for which we want to compute the correlator $\langle\sigma_i^z\sigma_j^z\rangle$.
    (b) Quantum circuit for the graph in (a). The dumbbells represent parameterized $\text{R}_{ZZ}$ rotations and the dark and light grey boxes are Hadamards and $R_X$ gates, respectively.
    The wire colors in (b) correspond to the color of the nodes in the graph in (a).
    }
    \label{fig:light_cone}
\end{figure}

For low-depth QAOA and sparse graphs, such as RR3, we can efficiently compute the expectation value $\langle H_C\rangle=\sum_{(i,j)\in E}\langle\sigma_i^z\sigma_j^z\rangle$ by considering the light-cone of each correlator $\sigma_i^z\sigma_j^z$.
Indeed, for depth-one QAOA each correlator $\sigma_i^z\sigma_j^z$ is only impacted by the gates applied to nodes in the direct neighborhood of $i$ and $j$, i.e., distance one nodes, see Fig.~\ref{fig:light_cone}.
For depth-two QAOA, we must consider all nodes that are at most at a distance of two away from $i$ and $j$ in $E$.
Therefore, to compute $\langle H_C\rangle$ for depth-two QAOA we create $|E|$ circuits each with at most 14 nodes.
In the circuit corresponding to $\langle\sigma_i^z\sigma_j^z\rangle$ we only measure the qubits that map to nodes $i$ and $j$, see Fig.~\ref{fig:light_cone}(b).

\section{Model comparison\label{sec:ml_comp}}

\begin{figure}
    \centering
    \includegraphics[width=\columnwidth]{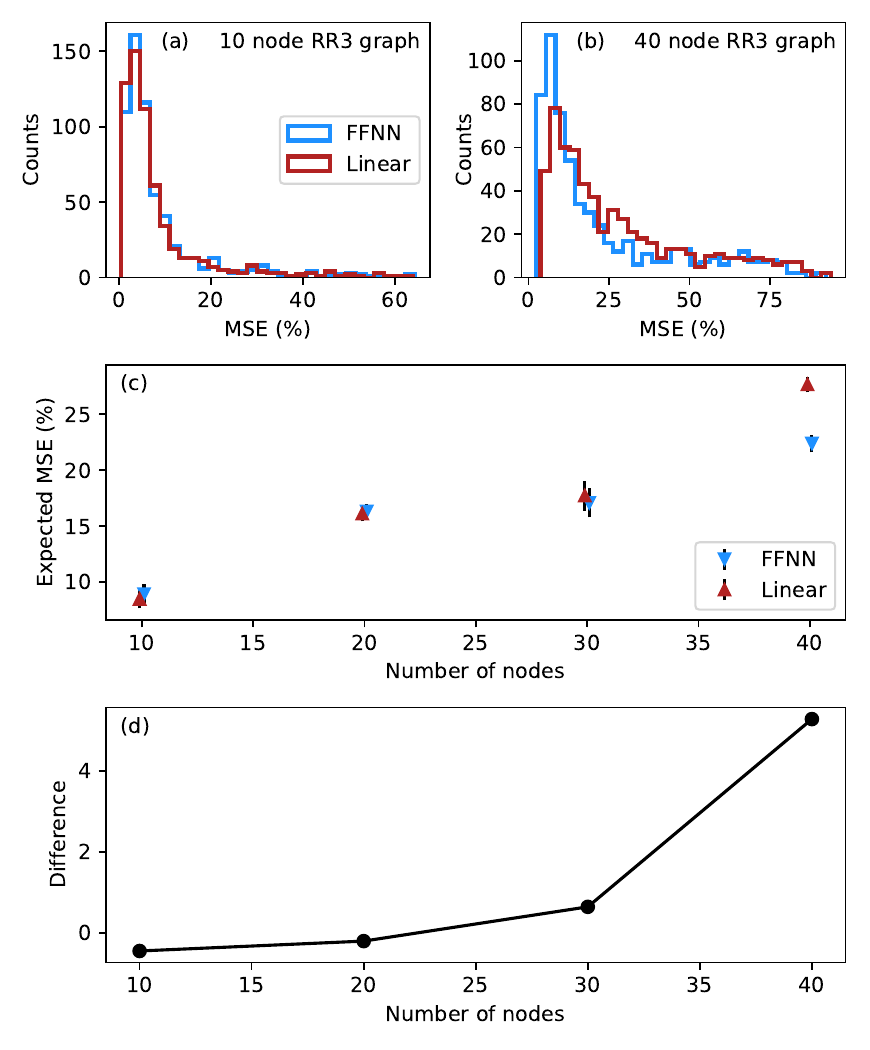}
    \caption{
    Comparison of a FFNN and a linear regression.
    (a) and (b) show the distribution of the MSE of a validation data set for the 10 and 40 node graphs, respectively.
    (c) MSE averaged over all ten validation data sets as a function of graph size.
    (d) Difference between the MSE of the FFNN and the linear regression.
    }
    \label{fig:ml_comp}
\end{figure}

Here, we compare the FFNN to a linear regression. 
The comparison is done on the hardware measured data presented in Fig.~\ref{fig:study} of the main text.
We resample the 3000 training data points ten times to generate ten training data sets made of 80\% of the data, i.e., 2400 sets of $|V|(|V|+1)/2$ expectation values, and 10 validation sets made of the remaining 20\% of the data. 
We train both a FFNN and a linear model on the ten data sets with 80\% of the data. 
Here, we employ the \texttt{MLPRegressor} and the \texttt{LinearRegression} from \texttt{sklearn}. 
Next, we compute the mean squared error between the $3|V|/2$ predicted $ZZ$ correlators and their ideal value. 
This results in a distribution of MSEs accross the 600 validation data points for each of the 10 validation data sets. We observe an increase in the MSE as the graph size increases. 
Furthermore, the MSE of the linear model increases faster than the MSE of the FFNN as the graph size increases, as exemplified by comparing Figs.~\ref{fig:ml_comp}(a) and (b) for one of the ten validation sets. 
Next, we compute the avarage over all MSEs at each graph size and the associated standard deviation of this mean, see Fig.~\ref{fig:ml_comp}(c) and their difference in (d). 
As the number of nodes in the graph increase the error of the FFNN becomes significantly lower than the error of the linear model. 
This test suggests that for the particular circuits that we employed the FFNN performs better than the linear model as the size of the quantum circuit increases.
This may be due to effects such as cross-talk and unitary gate errors, e.g., over- and under-rotations, that are not captured by a linear model like the depolarizing channel~\cite{czarnik2021error}.

\section{Additional hardware runs}

Here, we present depth-two QAOA data acquired on \emph{ibm\_nazca} and \emph{ibm\_kyiv} in addition to the data acquired on \emph{ibm\_brisbane}.
The data acquired on \emph{ibm\_nazca} was gathered under the same settings as the data on \emph{ibm\_brisbane}.
The data acquired on \emph{ibm\_kyiv} is produced with a smaller number of training circuits, i.e., 300 instead of 3000, and the hidden layer of the FFNN had 100 nodes for each graph size.
By contrast, the FFNN trained for \emph{ibm\_brisbane} and \emph{ibm\_nazca} had a number of hidden neurons equal to the average of the input and output number of neurons.

\begin{figure*}
    \centering
    \includegraphics[width=\textwidth]{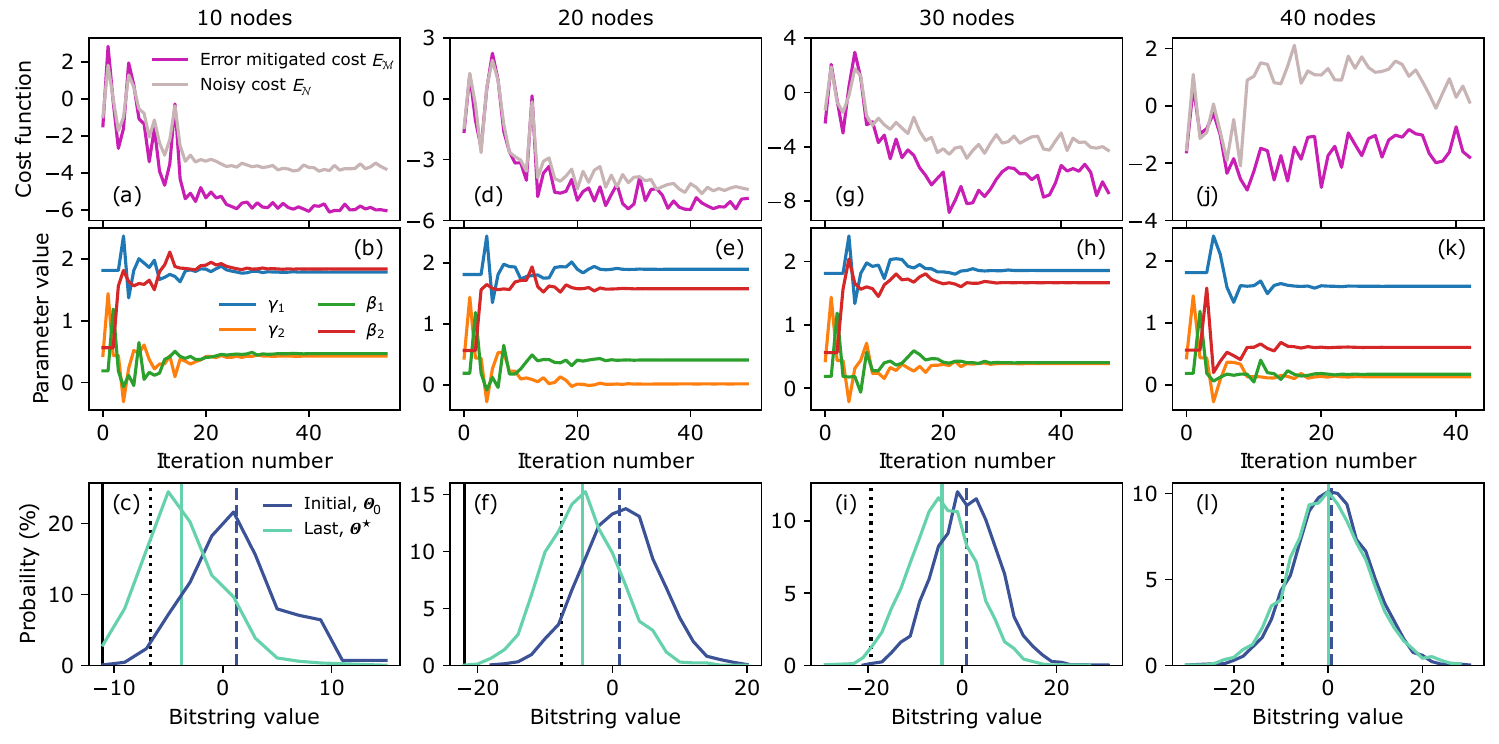}
    \caption{
    Run on \emph{ibm\_kyiv}.
    The training data was comprised of 300 circuits and the FFNN had one hidden layer with 100 nodes in all instances.
    The underlying graphs and other displayed quantities are identical to those in Fig.~\ref{fig:study} of the main text.
    }
    \label{fig:study_kyiv}
\end{figure*}

\begin{figure*}
    \centering
    \includegraphics[width=\textwidth]{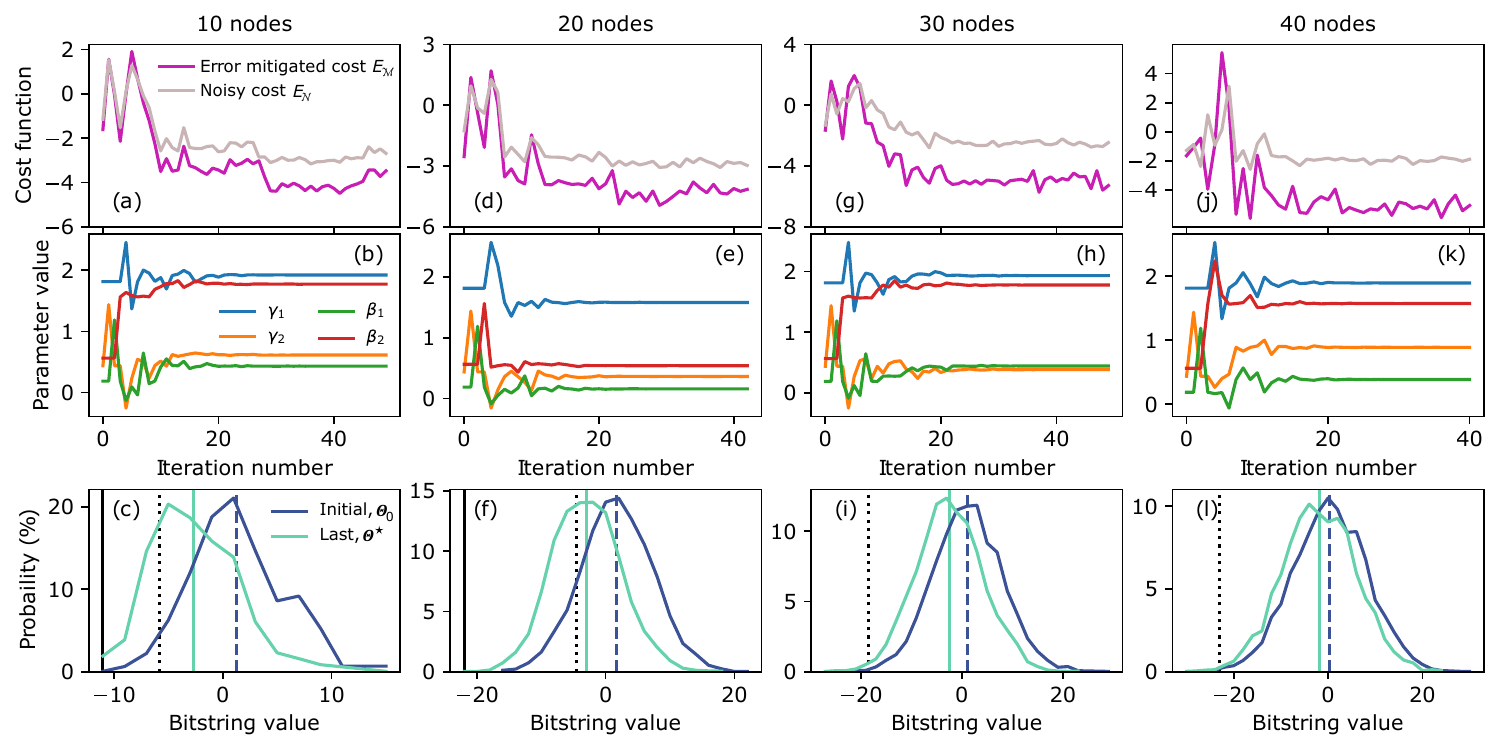}
    \caption{Run on \emph{ibm\_nazca}.
    The training data was comprised of 3000 circuits and the FFNN had one hidden layer with a number of nodes made of the average of the input and output.
    The underlying graphs and other displayed quantities are identical to those in Fig.~\ref{fig:study} of the main text.
    }
    \label{fig:study_nazca}
\end{figure*}

\clearpage

\bibliography{bibliography}

\end{document}